# Controlling Kondo-like Scattering at the SrTiO$_3$-based Interfaces

K. Han,[1,2] N. Palina,[1,3] S. W. Zeng,[1,2] Z. Huang,*,[1] C. J. Li,[1] W. X. Zhou,[1,2] D-Y. Wan,[1,2] L. C. Zhang,[1,2] X. Chi,[3] R. Guo,[1,4] J. S. Chen,[4] T. Venkatesan,[1,2,4,5,6] A. Rusydi,[1,2,3] and Ariando*,[1,2,6]

[1]NUSNNI-NanoCore, National University of Singapore, Singapore 117411, Singapore
[2]Department of Physics, National University of Singapore, Singapore 117542, Singapore
[3]Singapore Synchrotron Light Source, National University of Singapore, Singapore 117603, Singapore
[4]Department of Material Science & Engineering, National University of Singapore, Singapore 117575, Singapore
[5]Department of Electrical and Computer Engineering, National University of Singapore, Singapore 117576, Singapore
[6]National University of Singapore Graduate School for Integrative Sciences and Engineering (NGS), 28 Medical Drive, Singapore 117456, Singapore
*Correspondence email: nnihz@nus.edu.sg, ariando@nus.edu.sg.

**The observation of magnetic interaction at the interface between nonmagnetic oxides has attracted much attention in recent years. In this report, we show that the Kondo-like scattering at the SrTiO$_3$-based conducting interface is enhanced by increasing the lattice mismatch and growth oxygen pressure $P_{O2}$. For the 26-unit-cell LaAlO$_3$/SrTiO$_3$ (LAO/STO) interface with lattice mismatch being 3.0%, the Kondo-like scattering is observed when $P_{O2}$ is beyond 1 mTorr. By contrast, when the lattice mismatch is reduced to 1.0% at the (La$_{0.3}$Sr$_{0.7}$)(Al$_{0.65}$Ta$_{0.35}$)O$_3$/SrTiO$_3$ (LSAT/STO) interface, the metallic state is always preserved up to $P_{O2}$ of 100 mTorr. The data from Hall measurement and X-ray absorption near edge structure (XANES) spectroscopy reveal that the larger amount of localized Ti$^{3+}$ ions are formed at the LAO/STO interface compared to LSAT/STO. Those localized Ti$^{3+}$ ions with unpaired electrons can be spin-polarized to scatter mobile electrons, responsible for the Kondo-like scattering observed at the LAO/STO interface.**

## Introduction

Due to the discontinuity in degrees of freedom such as lattice, charge, spin and orbital, the interface can show unique property that cannot be found in bulk materials[1]. One well-known example is the LaAlO$_3$/SrTiO$_3$ (LAO/STO) interface[2], which can be tuned to exhibit conducting and magnetic properties, characterized by the two-dimensional electron gas (2DEG)[2-4] accompanied with Kondo effect[5] or ferromagnetic state[6-9]. This magnetic 2DEG has also been applied in electronic device such as magnetic tunnel junction[10]. Although this novel interface magnetism has been observed by different techniques[5-12], some contradicting results were found in recent studies. For examples, the magnetic moment density can vary from 10$^{-3}$ $\mu_B$[13] to 0.3 $\mu_B$ per unit cell[7] and persist up to room temperature[6,11]. The data from magnetotransport[5] and superconducting quantum interference device (SQUID)[6] showed that the magnetic interaction is enhanced by increasing an oxygen pressure during the sample growth, or reducing oxygen vacancy. However, the study on X-ray absorption spectroscopy (XAS) clearly demonstrated that formation of the oxygen vacancy is crucial for ferromagnetism at the interface[14]. In order to resolve these inconsistencies, the interplay between itinerant electrons and local magnetic moments is emphasized[15-19]. The different types of magnetic interaction can be established at the LAO/STO interface by changing parameters such as carrier density[17,18] and/or oxygen vacancy density[19], leading to the variation in magnetic moment density and contradicting results as reported[5-7,13,14]. In this scenario[15-19], the itinerant electrons can be provided by 2DEG. However, the origin of the local magnetic moment is still unclear and believed to be related to several factors such as localized electrons at Ti 3$d$ orbitals[9,14,15,18], cation antisite defect[20] and oxygen vacancies[21-23].

Hence, more information is needed for better understanding of the magnetic origin, especially the nature of localized electrons exhibiting local magnetic moments at the interface between nonmagnetic oxides.

Several studies have shown that the lattice-mismatch-induced interface strain can tune the electronic structure in the SrTiO$_3$-based systems, affecting the transport properties[24-26] and interface magnetism[27,28]. In order to investigate the influence of lattice mismatch on the observed magnetism at the interface, we compare the LAO/STO interface with another oxide interface (La$_{0.3}$Sr$_{0.7}$)(Al$_{0.65}$Ta$_{0.35}$)O$_3$/SrTiO$_3$ (LSAT/STO), which has been proved to be able to be conducting by Huang et al[29]. Both LAO and LSAT are band insulators which have bandgap larger than STO. The crystal structure of LAO and LSAT follows the perovskite ABO$_3$-type lattice, with the polar AO/BO$_2$ layer alternatively stacked along [100] axis. Moreover, both of them can induce 2DEG on STO substrate[29]. Considering the (pseudo)cubic lattice constant for LAO, LSAT and STO of 3.792 Å, 3.868 Å and 3.905 Å, respectively, the lattice mismatch is 3.0% at the LAO/STO interface and 1.0% at LSAT/STO.

In this paper, we present results from temperature-dependent and magnetic-field-dependent transport as well as X-ray absorption near edge structure (XANES) studies obtained for the LAO/STO and LSAT/STO interfaces. LAO and LSAT film are grown with various oxygen partial pressures $P_{O2}$ (0.05 – 5 mTorr). Our data show that when $P_{O2}$ is beyond 1 mTorr, the 26 unit cells (uc) LAO/STO interface begins to exhibit Kondo-like scattering, characterized by the resistance upturn (around 40 K) followed by the resistance saturation with negative isotropic magneto-resistance at low temperatures. In contrast, the LSAT/STO interfaces can always maintain the low-temperature metallicity when $P_{O2}$ is increased up to

100 mTorr. The XANES studies performed at Ti $L_{32}$-edge show that the $Ti^{3+}/Ti^{4+}$ ratio is larger at the LAO/STO interface, compared to the LSAT/STO interface. The $Ti^{3+}/Ti^{4+}$ ratio obtained from XANES should be regarded as a total number of electrons that occupy the Ti 3$d$ orbitals, including the itinerant and localized electrons. Considering the similar itinerant carrier density for LAO/STO and LSAT/STO interfaces, the larger amount of localized $Ti^{3+}$ ions the LAO/STO interface can be spin-polarized and scatter the mobile electrons, leading to the observed Kondo-like features.

**Results**

**Temperature-dependent transport property.** The data illustrating temperature-dependent sheet resistance $R_S$(T) for 26 uc LAO/STO and LSAT/STO interfaces with different $P_{O2}$ are shown in Figure 1. For the LAO/STO interface in Figure 1(a), the low-temperature metallic state ($dR_S/dT > 0$) is preserved in the samples with $P_{O2}$ below 1 mTorr, suggesting a normal 2DEG is established at the interface. When $P_{O2}$ is above 1 mTorr, the LAO/STO interfaces show a clear upturn of sheet resistance ($dR_S/dT < 0$) below 40 K, and $R_S$ becomes gradually saturated ($dR_S/dT \approx 0$) under further cooling. These features are different from a normal 2DEG with the low-temperature metallic state. However, for the 26 uc LSAT/STO interface in Figure 1(b), the metallic state of 2DEG can be always maintained down to 2 K when $P_{O2}$ is changing from 0.05 to 50 mTorr. Only a slight resistance upturn at ~ 15 K can be observed when $P_{O2}$ is increased to 100 mTorr. Hence, when increasing the $P_{O2}$, the metallic state is less favored at the interface with the lager lattice mismatch.

Usually, the upturn of sheet resistance is caused by either carrier scattering with low carrier mobility $\mu_S$, or carrier localization with low carrier density $n_S$. In Figure 2(a), all the LAO/STO interfaces show the decreasing $n_S$ from 100 to 2 K, probably due to the localization of the oxygen-vacancy-induced carriers as reported in the SrTiO$_{3-\delta}$[30,31]. However, carrier localization alone cannot explain resistance upturns in Figure 1(a). All the samples studied here, exhibit a similar low-temperature $n_S$ (2-3 ×10$^{13}$ cm$^{-2}$ at 2 K) independent on $P_{O2}$, and it is in contradiction with the resistance upturn, which has been observed only for the sample with high $P_{O2}$. On the other hand, for carrier mobility $\mu_S$ in Figure 2(b), the LAO/STO interfaces with high $P_{O2}$ (1 and 5 mTorr) show the decreasing $\mu_S$ below 40 K on cooling, while the increasing $\mu_S$ during cooling is observed at the interfaces with low $P_{O2}$ (0.05 – 0.5 mTorr). This data is consistent with the appearance of resistance upturn (metallic state) at the high-$P_{O2}$ (low-$P_{O2}$) interface. So, the upturns of $R_S(T)$ must be ascribed to the carrier scattering, instead of carrier localization.

Figure 2(c) presents the $n_S$ as a functional of temperature at the LSAT/STO interfaces. As can be seen the trend is similar to that of the LAO/STO samples. In particular, if compared with the LAO/STO interface, the LSAT/STO interfaces prepared at $P_{O2}$ = 0.05−0.5 mTorr exhibit a larger decrease of carrier density at low temperature, indicating the low $P_{O2}$ could create more oxygen vacancies at the LSAT/STO interface than at the LAO/STO. The carrier mobility $\mu_S$ of all the LSAT/STO samples is increasing under cooling, as shown in Figure 2(d). At the interface with a small lattice mismatch, e.g. LSAT/STO, $\mu_S$ is always higher as compared to the interface with a large lattice mismatch, e.g. LAO/STO. Moreover, even though the carrier density $n_S$ at 2 K is almost independent of $P_{O2}$, the carrier mobility $\mu_S$ at 2 K is very sensitive to $P_{O2}$ at both LAO/STO and LSAT/STO samples. As shown in Figure 2(e), when $P_{O2}$ is increased from 0.05 to 5 mTorr, $\mu_S$ at 2 K is reduced by factor of 300 for the LAO/STO (from 1,000 to 3 cm$^2$V$^{-1}$s$^{-1}$) and 15 for the LSAT/STO (from 23,000 to 1,500 cm$^2$V$^{-1}$s$^{-1}$) interface, respectively. For the low-$P_{O2}$ interfaces, the high carrier mobility might be due to the oxygen vacancy formation in the STO bulk. Comparison of the LAO/STO and LSAT/STO interfaces properties at different $P_{O2}$, reveals that: 1) the resistance upturn is caused by carrier scattering with low $\mu_S$, and 2) $\mu_S$ is more sensitive to $P_{O2}$ at the interface with a larger lattice mismatch.

**Magnetotransport property.** In Figure 3, the low-temperature (T= 2 K) magneto-resistance, defined by MR= [R(H)-R(0)]/R(0), is shown for the LAO/STO and LSAT/STO interfaces with different $P_{O2}$. The positive MR is observed in all the samples except the LAO/STO sample with $P_{O2}$= 5 mTorr. The LSAT/STO interface always exhibits the larger positive MR than LAO/STO interface with the same $P_{O2}$. For both interfaces, the magnitude of positive MR is consistently reduced with increasing $P_{O2}$ value. The positive MR is induced by the Lorentz-force-driven helical path for mobile carriers[32], and it can be enhanced by increasing $\mu_S$[33,34]. This is consistent with our observation that the larger positive MR appears at the higher mobility interface, of which the lattice mismatch is smaller and $P_{O2}$ is lower. However, the negative MR at the LAO/STO sample with $P_{O2}$= 5 mTorr is out of this picture, since the Lorentz force alone cannot induce the negative MR. The strong spin-orbit coupling may induce the large negative MR[35], but it does not correlate with the $R_S(T)$ data for the LAO/STO interface with $P_{O2}$= 5 mTorr. Two reasonable mechanisms can be proposed to explain the upturn in $R_S(T)$ and negative MR – one is the spin-related Kondo scattering[5,15,36], and the other is orbital-related weak anti-localization[37-39].

In order to distinguish these two different mechanisms, the MR curves with different field orientations are shown in Figure 4(a) for the LAO/STO interface with $P_{O2}$= 5 mTorr. The sample exhibits no observable difference in MR curves with changing the field orientation, and MR (*H*= 9 T) is always negative. This isotropic and negative MR confirms the spin-related Kondo-like scattering for the resistance upturn[5,15,36]. On the other hand, the metallic LSAT/STO interface exhibits the clear anisotropic MR, as shown in Figure 4(b). The positive MR is gradually suppressed by increasing the angle θ between the sample normal and field direction. Moreover, the negative MR appears when the in-plane field (θ= 90°) is applied. The angular-dependent MR, which is defined by AMR= [R(θ)-R(90°)]/R(90°) in Figure 4(c), clearly presents the isotropic MR at the 5 mTorr LAO/STO interface, medium anisotropic MR at the 0.05 mTorr LAO/STO and 5 mTorr LSAT/STO interface, and strong anisotropic MR at the 0.05 mTorr LSAT/STO interface. This suggests the AMR can be enhanced by lowering $P_{O2}$ and reducing lattice mismatch.

**X-Ray absorption near edge structure (XANES).** In order to clarify the origin of the Kondo-like scattering at the LAO/STO interface, Ti $L_{32}$-edge XANES spectra are compared in Figure 5(a) for TiO$_2$-terminated STO substrate (*t*-STO, reference for substrate), Ti$_2$O$_3$ (reference for Ti$^{3+}$), 10 uc LAO/STO and LSAT/STO interfaces with $P_{O2}$= 5 mTorr. The XANES is a powerful tool to examine the low-density Ti$^{3+}$ ions under a strong Ti$^{4+}$ background[40]. As can be seen, the LAO/STO interface exhibits a higher intensity around Ti$^{3+}$ states (see reference Ti$_2$O$_3$ spectrum) peaks denoted by red dash line as compared with LSAT/STO. Moreover, a linear combination fit analysis based on *t*-STO and Ti$_2$O$_3$ reference spectra revealed a Ti$^{3+}$/Ti$^{4+}$ ratio in a range of ~ 10% for the LAO/STO interface. In contrast, linear combination fit analysis for the LSAT/STO sample results in a negligible Ti$^{3+}$/Ti$^{4+}$ ratio of about 1% which is below the uncertainty range of XANES. Here we want to stress that $Ti^{3+}/Ti^{4+}$ ratio obtained from XANES should be proportional to the total number of electrons that occupy the Ti 3$d$ orbitals, including the mobile 2DEG and the localized $Ti^{3+}$ ions. Given that both interfaces show similar $n_S$ (3–4 ×$10^{13}$ cm$^{-2}$ from Hall measurement) of mobile 2DEG at room temperature, the larger amount of localized $Ti^{3+}$ ions is expected at the LAO/STO interface. One localized $Ti^{3+}$ ion can provide one unpaired electron, which can be spin-polarized and provide the local magnetic moment to scatter the mobile 2DEG at low temperatures, leading to the Kondo-like scattering at the LAO/STO interface.

**Discussions**

Our transport data demonstrate that the Kondo-like scattering is induced at the LAO/STO interface with high $P_{O2}$, but not at the LSAT/STO interface. The XANES analysis and Hall measurement identify a large amount of localized $Ti^{3+}$ ions at the LAO/STO interface, where the itinerant 2DEG can be scattered by the localized $Ti^{3+}$ ions with local magnetic moments to show the Kondo-like effect. However, there are still two questions needed to be addressed in our discussion. The first is why there are more localized $Ti^{3+}$ ions at the LAO/STO interface; the second is why $P_{O2}$ can influence the Kondo-like scattering.

For the first question, the different lattice mismatch at the LAO/STO and LSAT/STO interfaces is emphasized. As well documented, most of the localized electrons are located near the interface, where the interface disorders can lift the mobility edge for Anderson localization[41-44]. However, the interface disorders such as cation antisite defect[20] and oxygen vacancies[21-23] that may induce local magnetic moment should be at the same level for both interfaces, because the LAO/STO and LSAT/STO interfaces were fabricated under the same condition including laser energy, growth temperature and oxygen pressure. By contrast, the interface lattice distortion, especially for the STO layer that is close to the interface, must be much larger at the LAO/STO interface than the LSAT/STO interface due to the larger lattice mismatch and symmetry breaking at the LAO/STO interface. Such lattice distortions including the tetragonal-like $TiO_6$ deformation[25] and octahedral tilting[27,45] would narrow the Ti 3$d$ band, resulting in electron localization and magnetic interface[27,28]. Hence, when increasing the lattice mismatch from LSAT/STO to LAO/STO interface, the larger structural distortion is expected to produce more localized $Ti^{3+}$ ions and stronger Kondo-like scattering.

Regarding the influence of $P_{O2}$, calculations have shown that the Kondo effect is observable with low density of oxygen vacancy (high $P_{O2}$), if the oxygen vacancy interacting with Ti 3$d$ orbitals it can induce local magnetic moments[19,21,22]. Here, we argue that $P_{O2}$ can also tune the location of itinerant electrons, resulting in a stronger Kondo-like scattering for the higher $P_{O2}$. When $P_{O2}$ is low, not only the interface but also the bulk region of the STO substrate become conducting due to the oxygen vacancy. In this case, the conductive bulk region of STO could weaken the confinement potential of the interface electrons, so the itinerant electrons can travel away from the interface where the localized $Ti^{3+}$ ions are located[41-44], leading to a weaker magnetic scattering. Therefore, the mobile electrons are spatially separated from the localized $Ti^{3+}$ as shown in Figure 5(b), and the spin-relate Kondo-like scattering from localized $Ti^{3+}$ is very weak. When $P_{O2}$ is increasing, the propagation depth of mobile carriers in the STO substrate is greatly reduced[46]. In other words, by increasing $P_{O2}$ the mobile electrons are pushed to the interface with a better confinement[47]. So, as schematically shown in Figure 5(c), the itinerant electrons are much closer to the localized $Ti^{3+}$ ions and the stronger interaction between itinerant carriers and localized $Ti^{3+}$ are expected. It leads to the Kondo-like features, including resistance upturn and saturation, low carrier mobility, and isotropic negative MR, which are observed at the sample with increasing $P_{O2}$. This model can also explain the AMR behavior at the metallic interface as shown in Figure 4(b). By applying an in-plane magnetic field, the Lorentz force will drive the mobile carriers along the 2DEG normal to interact with the localized $Ti^{3+}$ close to the interface. So, when the magnetic field is changed from out-of-plane to in-plane ($\theta$ from $0^O$ to $90^O$), the spin-relate scattering arising from the localized $Ti^{3+}$ begins to take effect to suppress the positive MR and eventually show the negative in-plane MR.

## Conclusions

In summary, the crucial roles of lattice mismatch and growth oxygen pressure in Kondo-like effect has been demonstrated by comparing LAO/STO and LSAT/STO interfaces. For the LAO/STO interface with 3.0% lattice mismatch, the Kondo-like effect appears in the 26 uc sample when $P_{O2}$ is above 1 mTorr. For the LSAT/STO interface with 1.0% lattice mismatch, the metallic state is always preserved up to $P_{O2}$ of 100 mTorr. From the XANES and Hall measurement, a larger amount of the localized $Ti^{3+}$ is identified at the LAO/STO interface compared to the LSAT/STO interface. Those localized $Ti^{3+}$ ions can be spin-polarized and scatter the mobile electrons, leading to the observed Kondo-like features. Our results demonstrate that the Kondo-like effect at the $SrTiO_3$-based interface can be dually-controlled by lattice mismatch and $P_{O2}$, paving the path for engineering the interface magnetism at the functional oxide heterostructures.

## Methods

**Sample fabrication.** 26 unit cells (uc) of LAO and LSAT layers were deposited onto a TiO$_2$-terminated STO (001) substrates by pulsed laser deposition using a KrF laser (λ= 248nm). The LAO and LSAT single crystal targets are used for deposition. During the deposition, the laser repetition is kept at 1 Hz, laser fluence at 1.8 J/cm$^2$, growth temperature at 760 °C, and $P_{O2}$ varies from 0.05 to 100 mTorr. The deposition is monitored by *in-situ* reflection high energy electron diffraction (RHEED), from which the growth rate of 22-24 seconds per unit cell can be seen.

**Magnetotransport measurements.** The Hall bar is patterned on samples for measuring the transport property. The length of bridge is 160 *μm*, and the width is 50 *μm*. The transport property measurements were conducted in Physical Property Measurement System (Quantum Design, PPMS).

**X-Ray absorption near edge structure (XANES) measurements.** The XANES data have been recorded for the 10 uc LAO/STO and LSAT/STO interface with $P_{O2}$ = 5 mTorr. The thickness is chosen at 10 uc to guarantee the access to the interface during XANES measurements at the Ti $L_{32}$-edge. The XANES spectra were collected at the SINS beam-line at the Singapore Synchrotron Light Source (SSLS). To avoid possible contamination and surface modification, experiments were performed in UHV chamber with a background pressure of about 2x10$^{-10}$ mbar. All XANES spectra presented here were recorded *ex-situ* and at X-ray incident angle of 90$^O$ using total electron yield (TEY) mode.

## Acknowledgments

This work is supported by the MOE Tier 1 (Grant No. R-144-000-364-112 and R-144-000-346-112) and Singapore National Research Foundation (NRF) under the Competitive Research Programs (CRP Award No. NRF-CRP8-2011-06 and CRP Award No. NRF-CRP10-2012-02).

## Author Contributions

Z.H. and A. designed the experiment. K.H. and Z. H. prepared the samples. S.W.Z and C.J.L prepared the Hall bar pattern. K.H. and W.X.Z. conducted the electrical measurements with the assistance from D.Y.W and L.C.Z. The XANES measurement was conducted by N.P., and analyzed by N.P, X.C. and A.R. Insight and expertise on physical mechanism were provided by Z.H. and A., and discussed with T.V., R.G. and J.S.C. The manuscript was prepared by Z.H. and A., and fully revised by the other authors. The project was led by A.

## Additional Information

The authors declare no competing financial interests.

# References

1. Hwang, H. Y. *et al*. Emergent phenomena at oxide interfaces. *Nat. Mater.* **11**, 103–113 (2012).

2. Ohtomo, A. & Hwang, H. Y. A high-mobility electron gas at the LaAlO$_3$/SrTiO$_3$ heterointerface. *Nature* **427**, 423–426 (2004).

3. Nakagawa, N., Hwang, H. Y. & Muller, D. A. Why some interfaces cannot be sharp. *Nat. Mater.* **5**, 204–209 (2006).

4. Thiel, S., Hammerl, G., Schmehl, A., Schneider, C. W. & Mannhart, J. Tunable Quasi-two-dimensional electron gases in oxide heterostructures. *Science* **313**, 1942–1945 (2006).

5. Brinkman, A. *et al*. Magnetic effects at the interface between non-magnetic oxides. *Nat. Mater.* **6**, 493–496 (2007).

6. Ariando *et al*. Electronic phase separation at the LaAlO$_3$/SrTiO$_3$ interface. *Nat. Commun.* **2**, 188 (2011).

7. Li, L., Richter, C., Mannhart, J. & Ashoori, R. C. Coexistence of magnetic order and two-dimensional superconductivity at LaAlO$_3$/SrTiO$_3$ interfaces. *Nat. Phys.* **7**, 762–766 (2011).

8. Bert, J. A. *et al*. Direct imaging of the coexistence of ferromagnetism and superconductivity at the LaAlO$_3$/SrTiO$_3$ interface. *Nat. Phys.* **7**, 767–771 (2011).

9. Lee, J. S. *et al*. Titanium $d_{xy}$ ferromagnetism at the LaAlO$_3$/SrTiO$_3$ interface. *Nat. Mater.* **12**, 703–706 (2013).

10. Ngo, T. D. N. *et al*. Polarity-tunable magnetic tunnel junctions based on ferromagnetism at oxide heterointerfaces. *Nat. Commun.* **6**, 8035 (2015).

11. Bi, F. *et al*. Room-temperature electronically-controlled ferromagnetism at the LaAlO$_3$/SrTiO$_3$ interface. *Nat. Commun.* **5**, 5019 (2014).

12. Kalisky, B. *et al*. Critical thickness for ferromagnetism in LaAlO$_3$/SrTiO$_3$ heterostructures. *Nat. Commun.* **3**, 922 (2012).

13. Fitzsimmons, M. R. *et al*. Upper limit to magnetism in LaAlO$_3$/SrTiO$_3$ heterostructures. *Phys. Rev. Lett.* **107**, 217201 (2011).

14. Salluzzo, M. *et al*. Origin of interface magnetism in BiMnO$_3$/SrTiO$_3$ and LaAlO$_3$/SrTiO$_3$ heterostructures. *Phys. Rev. Lett.* **111**, 087204 (2013).

15. Lee, M., Williams, J. R., Zhang, S., Frisbie, C. D. & Goldhaber-Gordon, D. Electrolyte gate-controlled Kondo effect in SrTiO$_3$. *Phys. Rev. Lett*. **107**, 256601 (2011).

16. Banerjee, S., Erten, O. & Randeria, M. Ferromagnetic exchange, spin-orbit coupling and spiral magnetism at the LaAlO$_3$/SrTiO$_3$ interface. *Nat. Phys.* **9**, 626–630 (2013).

17. Joshua, A., Ruhman, J., Pecker, S., Altman, E. & Ilani, S. Gate-tunable polarized phase of two-dimensional electrons at the LaAlO$_3$/SrTiO$_3$ interface. *Proc. Natl. Acad. Sci. USA* **110**, 9633 (2013).

18. Ruhman, J., Joshua, A., Ilani, S. & Altman, E. Competition between Kondo screening and magnetism at the LaAlO$_3$/SrTiO$_3$ interface. *Phys. Rev. B* **90**, 125123 (2014).

19. Behrmann, M. & Lechermann, F. Interface exchange processes in LaAlO$_3$/SrTiO$_3$ induced by oxygen vacancies. *Phys. Rev. B* **92**, 125148 (2015).

20. Yu, L. & Zunger, A. A polarity-induced defect mechanism for conductivity and magnetism at polar-nonpolar oxide interfaces. *Nat. Commun.* **5**, 5118 (2014).

21. Lin, C. & Demkov, A. A. Electron correlation in oxygen vacancy in $SrTiO_3$. *Phys. Rev. Lett.* **111**, 217601 (2013).

22. Lin, C. & Demkov, A. A. Consequences of oxygen-vacancy correlations at the $SrTiO_3$ interface. *Phys. Rev. Lett.* **113**, 157602 (2014).

23. Pavlenko, N., Kopp, T. & Mannhart, J. Emerging magnetism and electronic phase separation at titanate interfaces. *Phys. Rev. B* **88**, 201104(R) (2013).

24. Berger, R. F., Fennie, C. J. & Neaton, J. B. Band gap and edge engineering via ferroic distortion and anisotropic strain: The case of $SrTiO_3$. *Phys. Rev. Lett.* **107**, 146804 (2011).

25. Huang, Z. *et al*. Biaxial strain-induced transport property changes in atomically tailored $SrTiO_3$-based systems. *Phys. Rev. B* **90**, 125156 (2014).

26. Bark, C. W. *et al*. Tailoring a two-dimensional electron gas at the $LaAlO_3/SrTiO_3$ (001) interface by epitaxial strain. *Proc. Natl. Acad. Sci. USA*. **108**, 4720 (2011).

27. Ganguli, N. & Kelly, P. J. Tuning ferromagnetism at interfaces between insulating perovskite oxides. *Phys. Rev. Lett.* **113**, 127201 (2014).

28. Nazir, S., Behtash, M. & Yang, K. The role of uniaxial strain in tailoring the interfacial properties of $LaAlO_3/SrTiO_3$ heterostructure. *RSC Adv.* **5**, 15682 (2015).

29. Huang, Z. *et al.* The effect of polar fluctuation and lattice mismatch on carrier mobility at oxide interfaces. *Nano Lett*. doi:10.1021/acs.nanolett.5b04814.

30. Liu, Z. Q. *et al*. Metal-insulator transition in $SrTiO_{3-x}$ thin films induced by frozen-out carriers. *Phys. Rev. Lett.* **107**, 146802 (2011).

31. Liu, Z. Q. *et al*. Origin of the two-dimensional electron gas at LaAlO$_3$/SrTiO$_3$ interfaces: the role of oxygen vacancies and electronic reconstruction. *Phys. Rev. X* **3**, 021010 (2013).

32. Wang, X. *et al*. Magnetoresistance of two-dimensional and three-dimensional electron gas in LaAlO$_3$/SrTiO$_3$ heterostructures: influence of magnetic ordering, interface scattering, and dimensionality. *Phys. Rev. B* **84**, 075312 (2011).

33. Bell, C., Harashima, S., Hikita, Y. & Hwang, H. Y. Thickness dependence of the mobility at the LaAlO$_3$/SrTiO$_3$ interface. *Appl. Phys. Lett.* **94**, 222111 (2009).

34. *Semiconductor Material and Device Characterization*, edited by D. K. Schroder (Wiley, New Jersey, 2006), p. 479.

35. Diez, M. *et al*. Giant negative magnetoresistance driven by spin-orbit coupling at the LaAlO$_3$/SrTiO$_3$ interface. *Phys. Rev. Lett.* **115**, 016803 (2015).

36. Das, S. *et al*. Kondo scattering in δ-doped LaTiO$_3$/SrTiO$_3$ interfaces: Renormalization by spin-orbit interactions. *Phys. Rev. B* **90**, 081107(R) (2014).

37. Hikami, S., Larkin, A. I. & Nagaoka, Y. Spin-Orbit Interaction and magnetoresistance in the two dimensional random system. *Progr. Theor. Phys.* **63**, 707 (1980).

38. Caviglia, A. D. et al. Tunable Rashba spin-orbit interaction at oxide interfaces. *Phys. Rev. Lett.* **104**, 126803 (2010).

39. Bergmann, G. Weak anti-localization – An experimental proof for the destructive interference of rotated spin ½. *Solid State Commun.* **42**, 815–817 (1982).

40. Chen, C., Avila, J., Frantzeskakis, E., Levy, A. & Asensio, M. C. Observation of a two-dimensional liquid of Fröhlich polarons at the bare SrTiO$_3$ surface. *Nat. Commun.* **6**, 8585 (2015).

41. Popović, Z. S., Satpathy, S. & Martin, R. M. Origin of the two-dimensional electron gas carrier density at the LaAlO$_3$ on SrTiO$_3$ interface. *Phys. Rev. Lett.* **101**, 256801 (2008).

42. Delugas, P. *et al*. Spontaneous 2-dimensional carrier confinement at the n-type SrTiO$_3$/LaAlO$_3$ interface. *Phys. Rev. Lett.* **106**, 166807 (2011).

43. Huang, Z. *et al*. Conducting channel at the LaAlO$_3$/SrTiO$_3$ interface. *Phys. Rev. B* **88**, 161107(R) (2013).

44. Pallecchi, I. *et al*. Giant oscillating thermopower at oxide interfaces. *Nat. Commun.* **6**, 6678 (2015).

45. Moon, S. Y. *et al*. Tunable conductivity at LaAlO$_3$/Sr$_x$Ca$_{1-x}$TiO$_3$ (0 ≤ x ≤ 1) heterointerfaces. *Appl. Phys. Lett.* **102**, 012903 (2013).

46. Basletic, M. *et al.* Mapping the spatial distribution of charge carriers in LaAlO$_3$/SrTiO$_3$ heterostructures. *Nat. Mater.* **7**, 621-625 (2008).

47. Annadi, A. *et al.* Fourfold oscillation in anisotropic magnetoresistance and planar Hall effect at the LaAlO$_3$/SrTiO$_3$ heterointerfaces: Effect of carrier confinement and electric field on magnetic interactions. *Phys. Rev. B* **87**, 201102(R) (2013).

**Figure 1. Temperature dependence of sheet resistance $R_S(T)$.** (a) $R_s(T)$ curves for 26 uc LAO/STO with $P_{O2}$ from 0.05 to 5 mTorr. (b) $R_s(T)$ curves for 26 uc LAO/STO with $P_{O2}$ from 0.05 to 100 mTorr. Inset: the schematic view of 2DEG exists at the LAO/STO (left) and LSAT/STO (right) interfaces with different lattice mismatch.

**Figure 2. Carrier density $n_S$ and carrier mobility $\mu_S$.** (a) $n_S(T)$ and (b) $\mu_S(T)$ curves for 26 uc LAO/STO interface prepared with $P_{O2}$ from 0.05 to 5 mTorr. (c) $n_S(T)$ and (d) $\mu_S(T)$ curves for 26 uc LSAT/STO interface prepared with $P_{O2}$ from 0.05 to 5 mTorr. (e) $\mu_S$ as a function of $P_{O2}$ for both interfaces. The blue and orange lines are guides to the eye, representing LSAT/STO and LAO/STO interfaces, respectively.

**Figure 3. $P_{O2}$-dependent MR with field perpendicular to the 2DEG plane at 2 K.** The MR curves for (a) LAO/STO and (b) LSAT/STO interfaces with different $P_{O2}$. The inset shows the applied magnetic field $H$ perpendicular to the 2DEG plane.

**Figure 4. The AMR behavior for both interfaces with different $P_{O2}$.** The MR curves with different $\theta$ at 2 K for (a) LAO/STO and (b) LSAT/STO interfaces with $P_{O2}$= 5 mTorr. The angle $\theta$ lies between the magnetic field and the normal of the interface, as shown in the inset of Figure 4(a). (c) The AMR curves for both interfaces with $P_{O2}$= 0.05 and 5 mTorr.

**Figure 5. The X-ray absorption near edge structure (XANES) and possible 2DEG location with respect to localized $Ti^{3+}$ ions.** (a) XANES for $t$-STO and $Ti_2O_3$ for reference (top), 10 uc

LSAT/STO and LAO/STO interface with $P_{O2}$ = 5 mTorr (middle), and residual values of both interfaces after being fitted by linear combination between $t$-STO and $Ti_2O_3$ (bottom). The red dashed lines indicate the peak positions for $Ti^{3+}$. (b) The spatially-separated localized $Ti^{3+}$ ions and mobile 2DEG at the metallic interface, where the spin scattering from the localized $Ti^{3+}$ is weak. (c) The localized $Ti^{3+}$ ions overlap the mobile 2DEG, resulting in the strong spin scattering and Kondo-like interface.

Figure 1

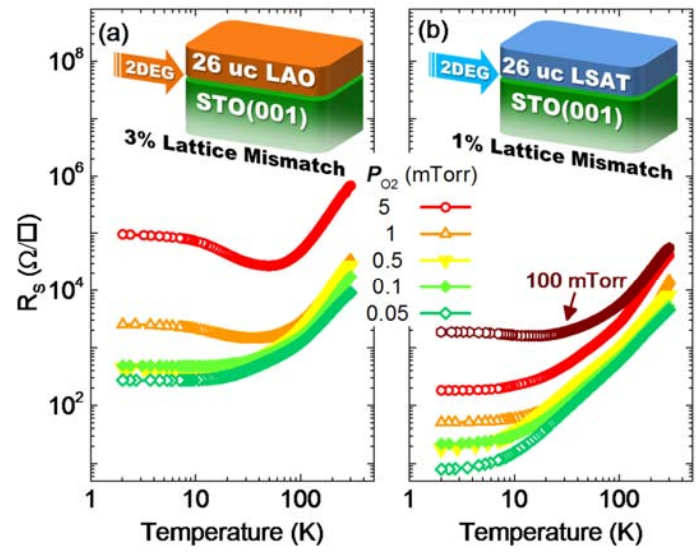

Figure 2

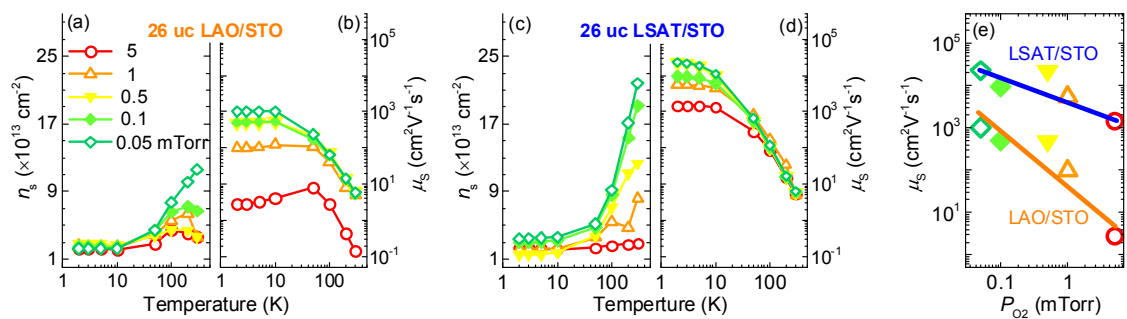

Figure 3

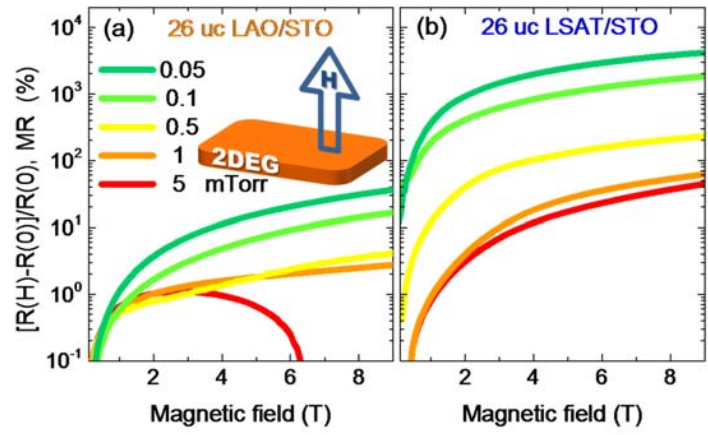

Figure 4

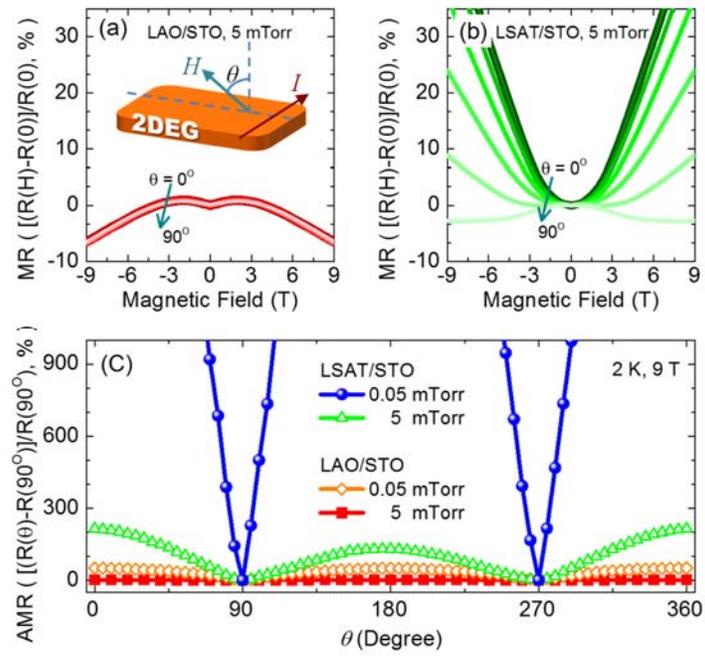

Figure 5

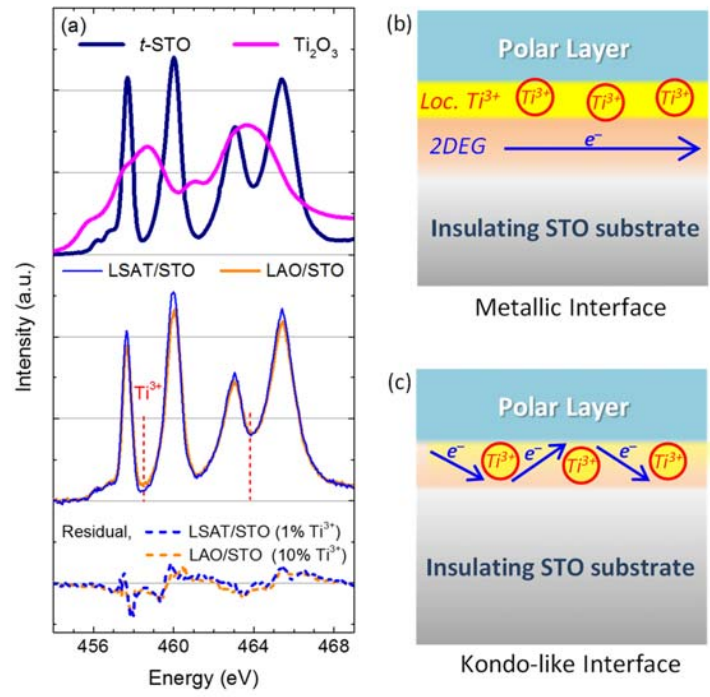